\documentclass[10pt,twocolumn]{article}
\usepackage{graphicx}
\usepackage[table]{xcolor}
\usepackage{ahsmeeting}
\usepackage[colorlinks=true]{hyperref}
\usepackage{amsmath}
\usepackage{amsfonts}
\usepackage{hyperref}
\usepackage{xcolor}

\hypersetup{
    colorlinks=true,
    linkcolor=blue,
    urlcolor=blue
}

\definecolor{halfgreen}{RGB}{0,128,0}
\definecolor{ahsred}{RGB}{192,0,0}

\newcommand{\forumyear}{2026}

\meeting{Presented at the Vertical Flight Society's 82nd Annual Forum~\&
\mbox{Technology} Display, West Palm Beach, FL, USA, May 5--7, 2026.
Copyright~\copyright~\forumyear\ by the Vertical Flight Society. All rights
reserved.}

\newcounter{isorefi}
{\endlist}

\begin{document}

\title{Adaptive Model Predictive Control of Nonlinear Generic Urban Air Mobility Using Linear Parameter-Varying Systems}

\author{
    \textbf{Tri Ngo} \\ 
    Research Scientist \\ 
    The University of Texas at Dallas \\
    Richardson, Texas, USA \\
    }
    \vspace{-1em}

\date{}

% Overriding the default maketitle behavior to inject the full-width link
% \twocolumn[
%   \begin{@twocolumnfalse}
%     \maketitle
    
%     \begin{center}
%         \vspace{-1.5em} % Adjust to control spacing between author and video
%         \textbf{Video:} \href{https://your-video-url-here.com}{Click here to watch the video}
%         \vspace{2em} % Adjust to control spacing between video and abstract/intro
%     \end{center}
%   \end{@twocolumnfalse}
% ]

\abstract{This paper presents an adaptive model predictive control (MPC) framework for nonlinear urban air mobility (UAM) vehicles operating across the full flight envelope. The proposed approach leverages a linear parameter-varying (LPV) representation to update the predictive model online, enabling accurate capture of strongly nonlinear and time-varying dynamics associated with distributed electric propulsion (DEP) eVTOL aircraft. To systematically address the high-dimensional and coupled nature of MPC tuning, a multi-objective evolutionary optimization strategy based on NSGA-II is employed, incorporating proper normalization of states and control inputs to ensure balanced weighting and meaningful exploration of the design space. The resulting controller explicitly accounts for actuator constraints and enables reconfigurable control allocation for fault-tolerant operation. The framework is evaluated in nonlinear simulations using NASA’s Generic Urban Air Mobility (GUAM) model and benchmarked against a robust servomechanism linear quadratic regulator (RSLQR). Results demonstrate that the proposed adaptive MPC achieves improved trajectory tracking and enhanced robustness under both nominal conditions and actuator degradation scenarios, including partial motor failure, while maintaining constraint satisfaction throughout all flight regimes.\textcolor{white}{asdfasdfasdfasdfasdfsadf}
 \textbf{Video:} \href{https://drive.google.com/file/d/1bmM0SYt7l28twWhZUEhw6EhSdjkgyiGa/view}{Click here to watch the video}
}

\maketitle

\section{Introduction}

The emergence of UAM represents a paradigm shift in metropolitan transportation, offering the potential to alleviate traffic congestion and enable rapid point-to-point connectivity through services such as airport shuttles, intercity connections, and short-distance air-taxi operations. The rapid development of electric vertical takeoff and landing (eVTOL) aircraft has positioned UAM as a transformative solution to future mobility challenges. These vehicles are expected to operate in dense urban environments under stringent safety, efficiency, and autonomy requirements, which place unprecedented demands on intelligent flight control systems. Unlike conventional aircraft, eVTOL configurations (e.g. tiltrotor, lift-plus-cruise) exhibit unconventional aerodynamics, strong nonlinearities across wide operating envelopes, and significant couplings between longitudinal and lateral dynamics, making controller design especially challenging. These hybrid configurations are particularly challenging due to their ability to transition between hover and forward flight, which induces dramatic variations in dynamic characteristics across the operating envelope. Thus, advanced control strategies are required to adapt effectively to changing flight regimes while preserving safety margins and mission performance. Moreover, UAM operations must contend with contingency events, strict safety constraints, and diverse performance requirements across operating conditions, underscoring the need for predictive, adaptive control methods \cite{gregory2020intelligent}

MPC has emerged as a powerful approach in aerospace applications due to its ability to handle multivariable dynamics, enforce actuator and state constraints, and optimize performance within a unified framework \cite{eren2017model}. Early studies have demonstrated its effectiveness in trajectory tracking, fault tolerance, and envelope protection for both fixed-wing and rotary-wing aircraft \cite{gopinathan1998multiple,ngo2016model,ngo2022variable, ngo2024robust,bauersfeld2021mpc,qu2023lpv, cavanini2021model, salzmann2023real}. However, its direct application to highly nonlinear systems such as eVTOL vehicles remains limited due to significant computational demands and strong dependence on accurate system models.
To address nonlinearities, gain-scheduling and linear parameter-varying (LPV) approaches are often employed in flight control, where local linear models are derived at multiple operating points and interpolated to approximate global dynamics \cite{cavanini2021model}. Such LPV models can be constructed using established system identification tools such as SIDPAC \cite{morelli2002system} or CIFER \cite{tischler2018system} from flight test data, as well as high-fidelity computational fluid dynamics (CFD) simulations. These methodologies are well established for fixed-wing and rotary-wing aircraft; however, their integration into LPV-based MPC frameworks for UAM systems remains relatively limited, with many existing studies relying on simplified dynamics and neglecting significant cross-axis coupling effects.
Another critical challenge is MPC parameter tuning, which has a strong influence on closed-loop performance. Conventional trial-and-error or heuristic tuning methods are often insufficient to balance competing objectives such as trajectory tracking accuracy, control effort, and robustness. In this context, multi-objective optimization techniques—particularly evolutionary algorithms such as genetic algorithms—provide a more systematic and scalable alternative, enabling improved performance across uncertain and highly nonlinear operating conditions.

This article proposes an adaptive MPC strategy for nonlinear UAM vehicles within an LPV framework. The key contributions are summarized as follows:
\begin{itemize}
\item Development of an adaptive MPC architecture that updates the predictive models online using an LPV representation. This enables effective handling of nonlinear and time-varying dynamics across the entire flight envelope. In addition, a systematic tuning framework based on a Multi-Objective Genetic Algorithm (MOGA) is employed to balance competing objectives, including tracking accuracy, control effort, robustness, and cross-axis coupling effects, which are particularly pronounced in distributed electric propulsion VTOL configurations.

\item Comprehensive evaluation of the proposed approach across multiple flight phases and contingency scenarios, including single electric motor failure cases, to assess robustness and resilience. Closed-loop simulations are conducted using NASA’s nonlinear Generic Urban Air Mobility (GUAM) model \cite{nasa_guam}. The results demonstrate that the proposed controller outperforms baseline methods, such as the Robust Servomechanism Linear Quadratic Regulator (RSLQR) \cite{cook2021robust}, in achieving core autonomous flight objectives while maintaining constraint satisfaction and ensuring fault-tolerant performance.
\end{itemize}

\section{NONLINEAR GENERIC URBAN AIR MOBILITY MODEL}
\subsection{Aircraft Dynamics}
\begin{figure}[ht] \begin{center}
\includegraphics[width=\columnwidth]{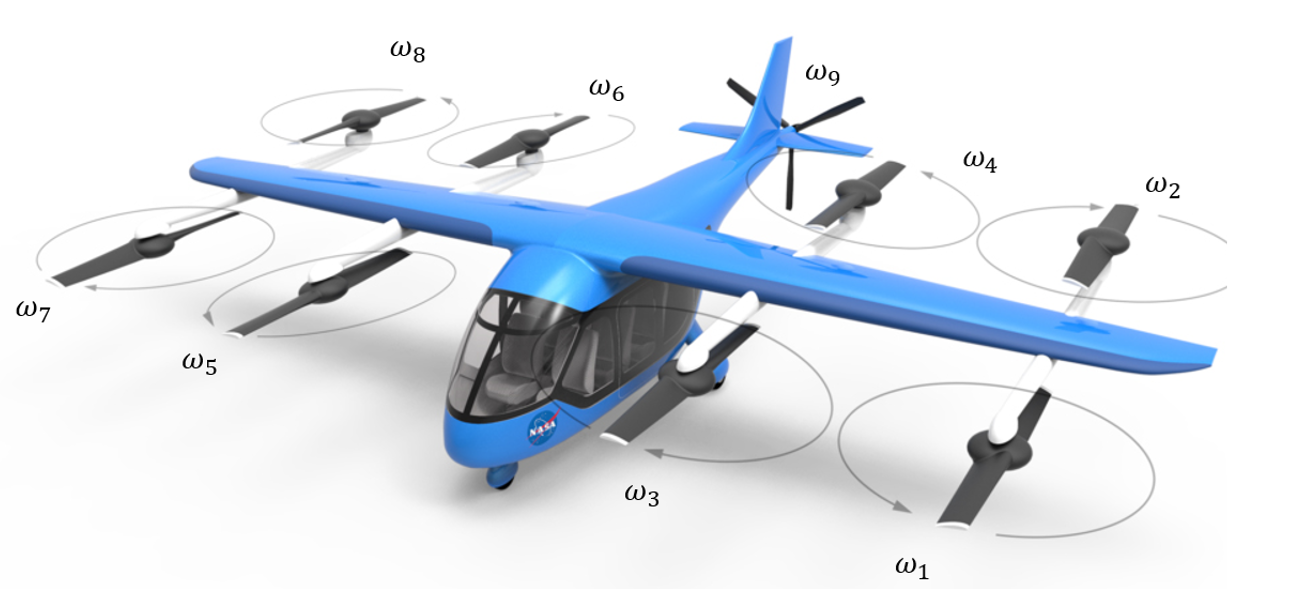}
\caption{Schematic of the NASA LPC concept \cite{simmons2021full}}\label{fig:LPC}
\end{center}\end{figure}

In this article, a control strategy is developed for a Lift+Cruise (LPC) eVTOL configuration. The vehicle, illustrated in Fig.~\ref{fig:LPC},  is a full-scale VTOL aircraft with distributed electric propulsion, consisting of eight fixed-pitch lift rotors and a rear-mounted, variable-pitch pusher propeller. The lift rotors are primarily utilized during takeoff, landing, and low-speed transition, as their operation incurs high energy consumption. In high-speed cruise, propulsion is provided exclusively by the pusher propeller, while the two-bladed lift rotors are aligned with the freestream to minimize aerodynamic drag.

The aircraft is represented as a six-degree-of-freedom (DoF) rigid body, with gyroscopic effects from the rotating propulsion components modeled as applied external moments. The translational and rotational dynamics are expressed in terms of forces X, Y, Z and moments L, M, N, which are derived from the identified aero-propulsive models. These models account for the combined contributions of the propulsion system, airframe aerodynamics, and their mutual interactions. The translational/rotational dynamics are given by:

\begin{align}
\dot{u} &= rv - qw - g \sin \theta + X/m \\
\dot{v} &= pw - ru + g \cos \theta \sin \phi + Y/m \\
\dot{w} &= qu - pv + g \cos \theta \cos \phi + Z/m \tag{1} \\
% \nonumber \\
I_x \dot{p} - I_{xz} \dot{r} &= L + (I_y - I_z)qr + I_{xz}pq - (\dot{h}_x + qh_z - rh_y) \\
I_y \dot{q} &= M + (I_z - I_x)pr + I_{xz}(r^2 - p^2) - (\dot{h}_y + rh_x - ph_z) \\
I_z \dot{r} - I_{xz} \dot{p} &= N + (I_x - I_y)pq - I_{xz}qr - (\dot{h}_z + ph_y - qh_x) \\
% \nonumber \\
\boldsymbol{\dot{\eta}} = S\boldsymbol{\omega}, \quad S &= 
\begin{bmatrix}
1 & \sin\phi \tan\theta & \cos\phi \tan\theta \\
0 & \cos\phi & -\sin\phi \\
0 & \sin\phi/\cos\theta & \cos\phi/\cos\theta
\end{bmatrix}
\end{align}

\subsection{Full-Envelope Aero-Propulsive Model}

In \cite{simmons2021full}, an aero-propulsive modeling methodology for a lift-plus-cruise eVTOL configuration is presented, leveraging design of experiment techniques and CFD simulations. The modeling framework covers the full operational envelope, encompassing four distinct flight regimes: hover, transition, cruise, and glide. Model adequacy is evaluated using validation data independent from the identification dataset, demonstrating sufficient predictive accuracy across regimes. The Rapid Aero Modeling (RAM) process is specifically tailored to capture lift-plus-cruise vehicle attributes and regime-specific dynamics. Nonlinearities inherent in eVTOL aerodynamics and propulsion are represented using multiple polynomial fits. A sample of a model for the propeller thrust coefficient quadratic in advance ratio is given as below. 

% C_{T} = C_{T_{o}} + C_{T_{J}}J + C_{T_{J^2}}J^{2} = C_{T_{o}} + C_{T_{J}}\left(\frac{V}{nD}\right) + C_{T_{J^2}}\left(\frac{V}{nD}\right)^{2} 

\begin{equation}
    C_T = C_{T_0} + C_{T_J} J + C_{T_{J^2}} J^2 = C_{T_0} + C_{T_J} \left( \frac{V}{nD} \right) + C_{T_{J^2}} \left( \frac{V}{nD} \right)^2
    \label{eq:thrust_coeff}
\end{equation}

This coefficient model can be converted to the dimensional thrust ($T$) using the standard thrust coefficient definition, resulting in the following equation. 

% T = \rho n^{2} D^{4} \left[ C_{T_{o}} + C_{T_{J}} \left( \frac{V}{nD} \right) + C_{T_{J^{2}}} \left( \frac{V}{nD} \right)^{2} \right] = \rho D^{2} \left( C_{T_{o}} D^{2} n^{2} + C_{T_{J}} DnV + C_{T_{J^{2}}} V^{2} \right)

\begin{equation}
    \begin{aligned}
    T &= \rho n^2 D^4 \left[ C_{T_0} + C_{T_J} \left( \frac{V}{nD} \right) + C_{T_{J^2}} \left( \frac{V}{nD} \right)^2 \right] \\
      &= \rho D^2 \left( C_{T_0} D^2 n^2 + C_{T_J} DnV + C_{T_{J^2}} V^2 \right)
    \end{aligned}
    \label{eq:dimensional_thrust}
\end{equation}

The explanatory and response variables are tailored to each flight regime to capture the dominant physics. In hover and transition, propulsive effects dominate, so the models predict body-axis forces and moments ($X, Y, Z, L, M, N$) based on body velocities, control deflections, and propeller speeds. In cruise and glide, aerodynamic forces prevail, and the models predict nondimensional aerodynamic coefficients ($C_x, C_y, C_z, C_l, C_m, C_n$) as functions of airspeed, angle of attack, sideslip, and control inputs, with the cruise regime also including propulsor operating parameters. This regime-specific formulation improves fidelity while reducing modeling error across the full operational envelope. For further details, the interested reader is referred to Simmons et al. \cite{simmons2021full}.

\section{BASELINE LINEAR QUADRATIC REGULATOR (LQR) CONTROLLER}

In this study, a robust servo-mechanism linear quadratic regulator (RSLQR) structure from \cite{cook2021robust} is employed as the baseline longitudinal and lateral/directional controllers for comparison. The RSLQR augments the system dynamics with an additional state representing the integral of the tracking error, thereby embedding integral action within the optimal control framework as follows. 

\begin{equation}
    \begin{bmatrix}
        \dot{x}_i \\ \dot{x}
    \end{bmatrix} = 
    \begin{bmatrix}
        0 & C \\ 0 & A
    \end{bmatrix}
    \begin{bmatrix}
        x_i \\ x
    \end{bmatrix} + 
    \begin{bmatrix}
        D \\ B
    \end{bmatrix} u + 
    \begin{bmatrix}
        -I \\ 0
    \end{bmatrix} r
    \label{eq:augmented_state_space}
\end{equation}
where \( x_i \in \mathbb{R}^p \) denotes the integral of the tracking error, \( x \in \mathbb{R}^n \) is the system state vector, \( r \in \mathbb{R}^p \) represents the reference input, and \( u \in \mathbb{R}^m \) is the control input vector. The overall structure comprises two components: a servo-tracking controller for reference command following and a state-feedback controller for system stabilization.
The control system is designed using a set of specific trim points (equilibrium conditions) covering various flight speeds, climb rates, and turn rates. A gain-scheduled approach is adopted, wherein control gains of RSLQR are scheduled based on airspeed, climb rate, and turn rate, thereby ensuring consistent control performance and a unified set of control commands across all flight regimes.

\section{ADAPTIVE MODEL PREDICTIVE CONTROL (MPC) FRAMEWORK}

\begin{figure}[t] \begin{center}
\includegraphics[width=\columnwidth]{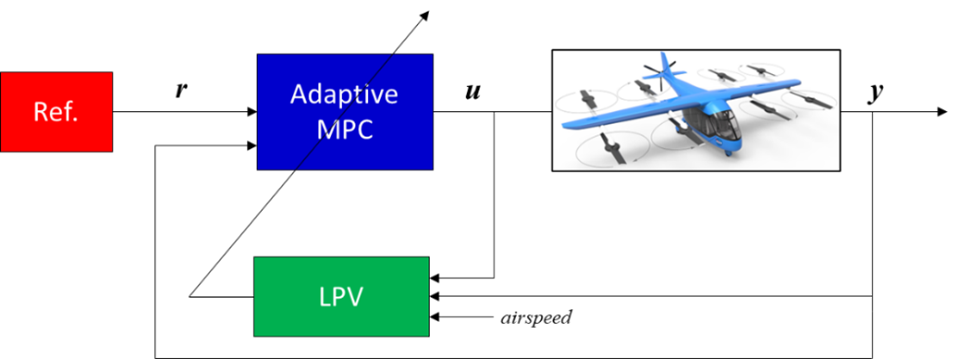}
\caption{Schematic of the adaptive MPC design}\label{fig:MPC_diagram}
\end{center}\end{figure}

Standard MPC relies on a linear time-invariant (LTI) model to predict the system’s future behavior \cite{maciejowski2007predictive}. However, when the plant exhibits strong nonlinearities or undergoes significant variations over time, the accuracy of LTI-based predictions can deteriorate, leading to unacceptable control performance. Adaptive MPC addresses this limitation by updating the prediction model to reflect changing operating conditions. While the overall model structure remains fixed, the model parameters are allowed to evolve with time. At the beginning of each control interval, the controller employs a LPV model that best represents the system dynamics under the current conditions (see Fig.~\ref{fig:MPC_diagram}). Following the standard MPC design, a cost function $V(k)$ is defined as:
\begin{equation}
\begin{aligned}
V(k) = \left\{ \sum_{i=0}^{H_p-1} \left( \sum_{j=1}^{p} w_{i+1,j}^y \left| (y_j(k+i+1|k) - r_j(k+i+1)) \right|^2 \right. \right. \\
\left. \left. + \sum_{j=1}^{m} w_{i+1,j}^u |u_j(k+i|k)|^2 + \sum_{j=1}^{m} w_{i+1,j}^{\Delta u} |\Delta u_j(k+i|k)|^2 \right) \right\}
\end{aligned}
\end{equation}

The MPC action at time step k is obtained by solving the following optimization problem.

\begin{equation}
\begin{aligned}
    & \underset{\Delta u(k|k), \dots, \Delta u(k+H_m-1|k)}{\text{minimize}} \quad V(k) \\
    \text{s.t.} \quad 
    & x(k+1) = A_d(p)\,x(k) + B_d(p)\,u(k) \\
    & y(k) = C_d(p)\,x(k) + D_d(p)\,u(k) \\
    & |\theta| \leq 30^\circ,\quad |\phi| \leq 30^\circ, \\
    & -30^\circ \leq \delta_f \leq 0, \\
    & -30^\circ \leq \delta_a,\ \delta_e,\ \delta_r \leq 30^\circ, \\
    & 0 \leq \omega_i \leq 168 \ \text{rad/s}, \quad i = 1,\dots,8, \\
    & 0 \leq \omega_9 \leq 209 \ \text{rad/s}
\end{aligned}
\label{eq:mpc_formulation}
\end{equation}
Eq.~(11) represents the LPV model of the eVTOL discretized with sampling time \( T_s \), and parameterized by the scheduling parameter $p$. The vectors \( x \), \( r \), \( u \), and \( \Delta u \) denote the system states, reference inputs, control inputs, and input increments, respectively. The output vector \( y \in \mathbb{R}^q \) corresponds to the controlled variables in the longitudinal and lateral/directional channels. The notation \( (\cdot)_j \) refers to the \( j \)-th component of a vector, while \( (k+i\,|\,k) \) denotes the prediction at time step \( k+i \) based on information available at time \( k \). The parameters \( q \), \( m \), \( H_p \), and \( H_m \) represent the number of outputs, number of inputs, prediction horizon, and control horizon, respectively. The weighting coefficients \( w^y \), \( w^u \), and \( w^{\Delta u} \) are nonnegative and are tuned to balance tracking performance, control effort, and actuator smoothness across different flight conditions.
The optimization problem is subject to both input and output constraints, including actuator saturation limits derived from the physical capabilities of the eVTOL system. In addition, output constraints are incorporated to accommodate large attitude excursions required during hover and low-speed flight, while maintaining acceptable levels of agility and passenger comfort.

\section{Tuning of MPC Parameters}

Careful tuning of MPC parameters is essential for achieving high-performance closed-loop behavior. Although manual tuning of weighting factors is commonly used in practice, it presents several challenges, including the high dimensionality of the parameter space, the strong nonlinear coupling inherent in eVTOL dynamics, and the interaction between multiple control channels governed by coupled MPC formulations. While acceptable performance may sometimes be obtained by tuning individual channels (e.g., longitudinal or lateral dynamics) independently, combining them often introduces cross-coupling effects that can degrade overall closed-loop performance and, in some cases, stability.

To address these challenges, conventional optimization-based tuning approaches formulate the selection of MPC weights as an optimization problem that minimizes a scalar performance index evaluated over closed-loop simulations. Typical methods include genetic algorithms, Bayesian optimization, and particle swarm optimization (PSO). However, when control objectives are inherently conflicting—such as trajectory tracking accuracy, control effort, and passenger comfort—a single aggregated cost function may not adequately capture the trade-offs between competing objectives. In such cases, multi-objective evolutionary algorithms provide a more systematic alternative. In particular, the Non-dominated Sorting Genetic Algorithm II (NSGA-II) \cite{deb2002fast} can be used to directly compute the Pareto front, enabling explicit visualization and selection of trade-offs among competing objectives.

It is important to note that, prior to applying optimization solvers, the objective functions must be properly defined and normalized, and appropriate bounds for the tuning parameters must be specified. Careful selection of these ranges is essential to ensure meaningful exploration of the parameter space and to avoid ill-conditioned or physically unrealistic solutions.
A critical step in the automated tuning framework is the normalization of both outputs and control inputs prior to optimization. In this work, normalization is achieved through scale factors assigned to each MPC variable, ensuring that all signals entering the objective functions have comparable magnitudes and preventing numerical bias toward states or actuators with larger physical units. An example of the MPC normalization parameters is provided in Table~\ref{tbl:mpc_scaling}.
% For the longitudinal controller, each output variable—forward and vertical velocities, pitch rate, and pitch angle—is normalized using dedicated scaling factors (``ScaleFactor''), which are selected to reflect their typical operating ranges and to increase sensitivity in key tracking variables such as velocity. Similarly, all manipulated variables, including the distributed rotor thrusts, pusher rotor, and control surfaces, are normalized using appropriate scale factors based on their physical limits and nominal operating conditions.

\begin{table}[htbp]
\centering
\caption{Normalization parameters}
\label{tbl:mpc_scaling}
\begin{tabular}{ll}
\hline \hline
\textbf{Parameter} & \textbf{Value / Range} \\
\hline
\multicolumn{2}{c}{\textit{Longitudinal Outputs}} \\
Forward velocity ($u$)  & 200 [ft/s] \\
Vertical velocity ($w$)  & 20 [ft/s] \\
Pitch rate ($q$)        & 30 [deg/s] \\
Pitch angle ($\theta$)  & 90 [deg] \\ \hline

\multicolumn{2}{c}{\textit{Lateral Outputs}} \\
Lateral velocity ($v$) & 20 [ft/s] \\
Roll rate ($p$)        & 100 [deg/s] \\
Yaw rate ($r$)         & 30 [deg/s] \\
Roll angle ($\phi$)    & 90 [deg] \\ \hline

\multicolumn{2}{c}{\textit{Control Inputs}} \\
Distributed rotors (N1--N8) & $[0,\;1600]$ [RPM] \\
Pusher rotor (N9)           & $[0,\;2000]$ [RPM] \\
Control surfaces (N10--N11) & $[-30,\;30]$ [deg] \\ \hline \hline
\end{tabular}
\end{table}

\vspace{-.1in}

\begin{table}[htbp]
\centering
\caption{Example of MPC tuning parameters}
\label{tbl:mpc_params}
\begin{tabular}{ll}
\hline \hline
\textbf{Parameter} & \textbf{Value} \\
\hline
Sampling time $T_s$ (s) & 0.05 \\ 
Prediction / control horizon $[H_p, H_c]$ & [50, 10] \\ \hline

\multicolumn{2}{c}{\textit{Output Weights}} \\
Longitudinal weights $[w_u, w_w, w_q, w_{\theta}]$ & [10, 5, 0.1, 5] \\
Lateral weights $[w_v, w_p, w_r, w_{\phi}]$ & [1, 10, 50, 0.001] \\ \hline

\multicolumn{2}{c}{\textit{Control Weights}} \\
Outer-loop gains $[k_p, k_x]$ & [0.1, 0.1] \\
Control surfaces $[k_a, k_{ra}]$ & [0.5, 10] \\
Propellers $[k_r, k_{rr}]$ & [0.5, 1] \\ \hline \hline
\end{tabular}
\end{table}

Table~\ref{tbl:mpc_params} summarizes key parameters used in the MPC formulation. The weighting coefficients \( w_u \), \( w_w \), \( w_q \), and \( w_{\theta} \) penalize tracking errors in the longitudinal states, while \( w_v \), \( w_p \), \( w_r \), and \( w_{\phi} \) penalize lateral tracking errors to promote smooth and stable motion. The additional tuning gains \( k_x \) and \( k_p \) shape the outer-loop command structure by mapping position errors to desired velocity commands and heading errors to yaw-rate commands, respectively. The gains \( k_a \), \( k_{ra} \), \( k_r \), and \( k_{rr} \) further adjust control allocation priorities and actuator coordination, including command magnitude limits and slew-rate penalties in the MPC cost function, enabling balanced utilization of distributed propulsion and aerodynamic control surfaces.

% $r(t) = \begin{bmatrix} \bar{\text{v}}_d(t) + k_{x}\bar{e}(t) \\ \psi_d + k_{p}e_{\psi}(t) \end{bmatrix}$

% Finally, actuator constraints define the feasible operating envelope of the system, including rotor speed limits for the distributed propellers (N1--N8 and N9) and control surface deflection limits, ensuring that all optimized control commands remain physically realizable.

% Trajectory
\begin{figure*}[htbp] \begin{center}
\includegraphics[width=\textwidth,trim={1.5cm 3cm 1.5cm 8cm}, clip]{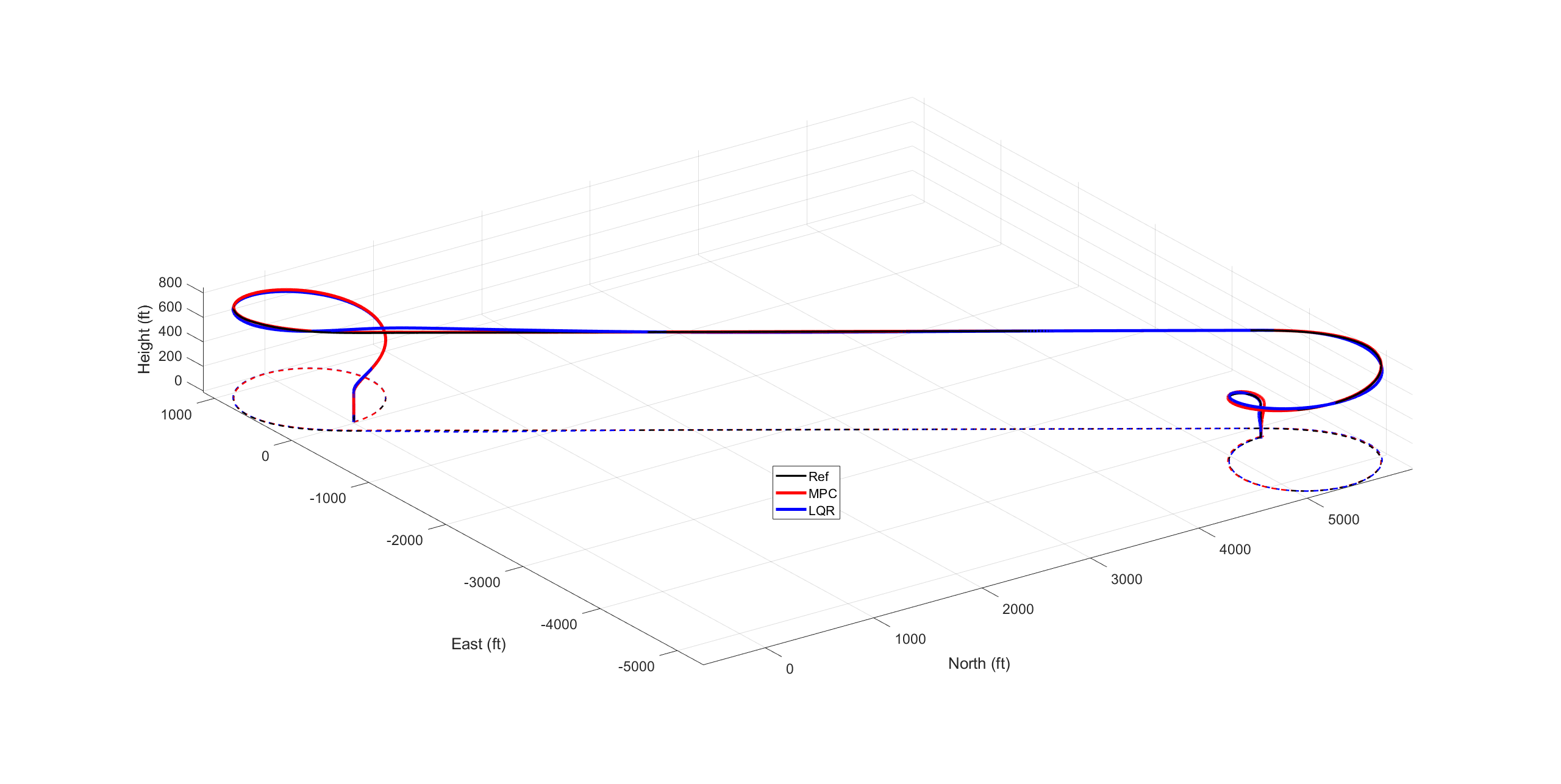}
\caption{Trajectories of eVTOL achieved using the LQR and MPC controllers}\label{fig:MPC_LQR_traj}
% \vspace{-0.5in}
\end{center}\end{figure*}

\section{Numerical Simulations and Performance Evaluation}

This section evaluates the proposed MPC architecture through nonlinear flight simulations of the NASA LPC eVTOL concept implemented in MATLAB/Simulink, covering the full operational envelope of the vehicle. The controller explicitly accounts for the nonlinear and time-varying dynamics of the aircraft through an LPV model formulation while optimally distributing control effort among the available effectors during all transition phases, ensuring stable and accurate trajectory tracking under both nominal and degraded conditions. The evaluation considers representative mission profiles, including hover, climb, turning, cruise, and descent maneuvers. Two scenarios are investigated: (i) nominal operation with all DEP actuators fully functional, and (ii) fault conditions in which one propeller experiences partial or complete loss of control authority, thereby highlighting the system’s fault-tolerant capabilities.

\subsection{All-regime trajectory tracking} 
The adaptive MPC framework is evaluated through simulations of an all-regime flight scenario, in which the aircraft tracks a prescribed trajectory presented in \cite{cook2021robust}. As shown in Fig.~\ref{fig:MPC_LQR_traj}, the maneuver sequence begins with a stationary hover, followed by a constant-radius helical ascent. The vehicle then executes a constant-radius helical ascent during the transition phase. It subsequently accelerates to a forward flight speed of 100 ft/s and maintains this cruise condition for a short duration. The trajectory concludes with a constant-radius helical descent, during which the aircraft decelerates and returns to a hover state.
The transition controller is implemented using the proposed adaptive MPC architecture described above, with RSLQR included for comparison. Both controllers employ linearized dynamic models at different equilibrium points, derived from steady, level-flight conditions as described in \cite{simmons2021full}.
 
% The control effector weighting matrices are determined via a genetic algorithm, prioritizing the use of propellers, elevons, and ruddervators as primary control effectors. A significantly higher penalty is assigned to flap deflections to minimize their use in control allocation.

\begin{figure}[htbp] % Remember: figure* usually prefers [t] over [b]
    \centering    \includegraphics[width=.8\columnwidth, trim={0cm 2cm 1cm 2cm}, clip]{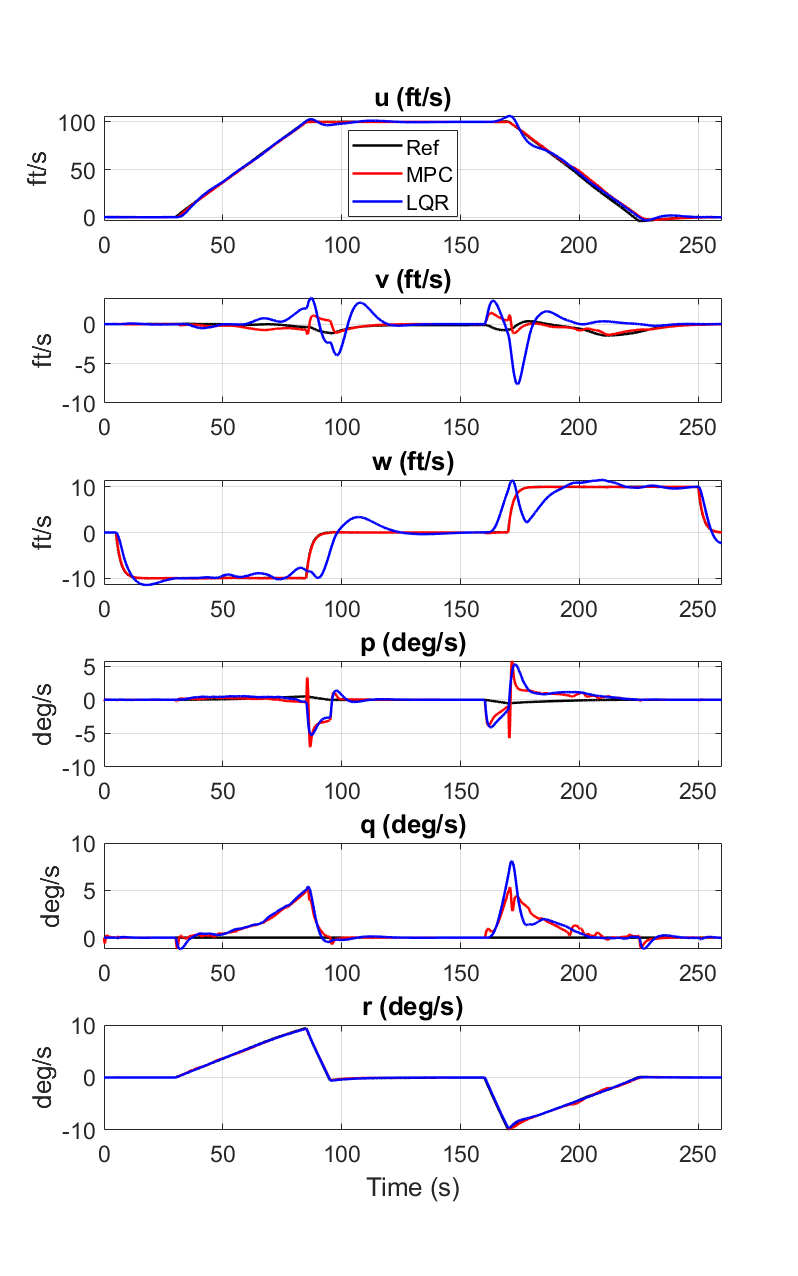}
    % \vspace{-1 in}
    \caption{Translational and rotational velocities.}
    \vspace{-.5in}
    \label{fig:MPC_LQR_vel}
\end{figure}

\begin{figure}[htbp] \begin{center}
\includegraphics[width=.8\columnwidth, trim={1cm 1cm 1cm .5cm}, clip]{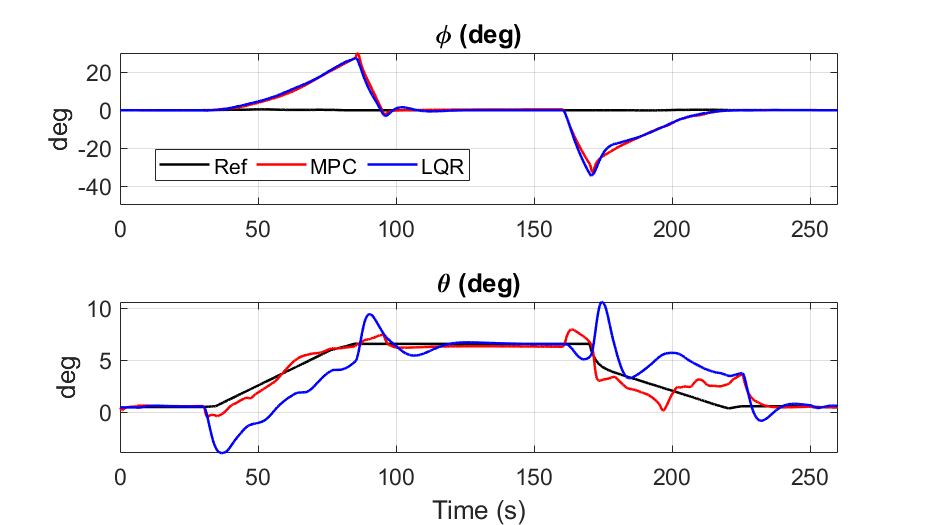}
\caption{Roll and pitch angles.}\label{fig:MPC_LQR_Euler}
\end{center}\end{figure}

\begin{figure}[htbp] \begin{center}
\includegraphics[width=.8\columnwidth,trim={1cm 1.5cm 1.5cm 1.5cm}, clip]{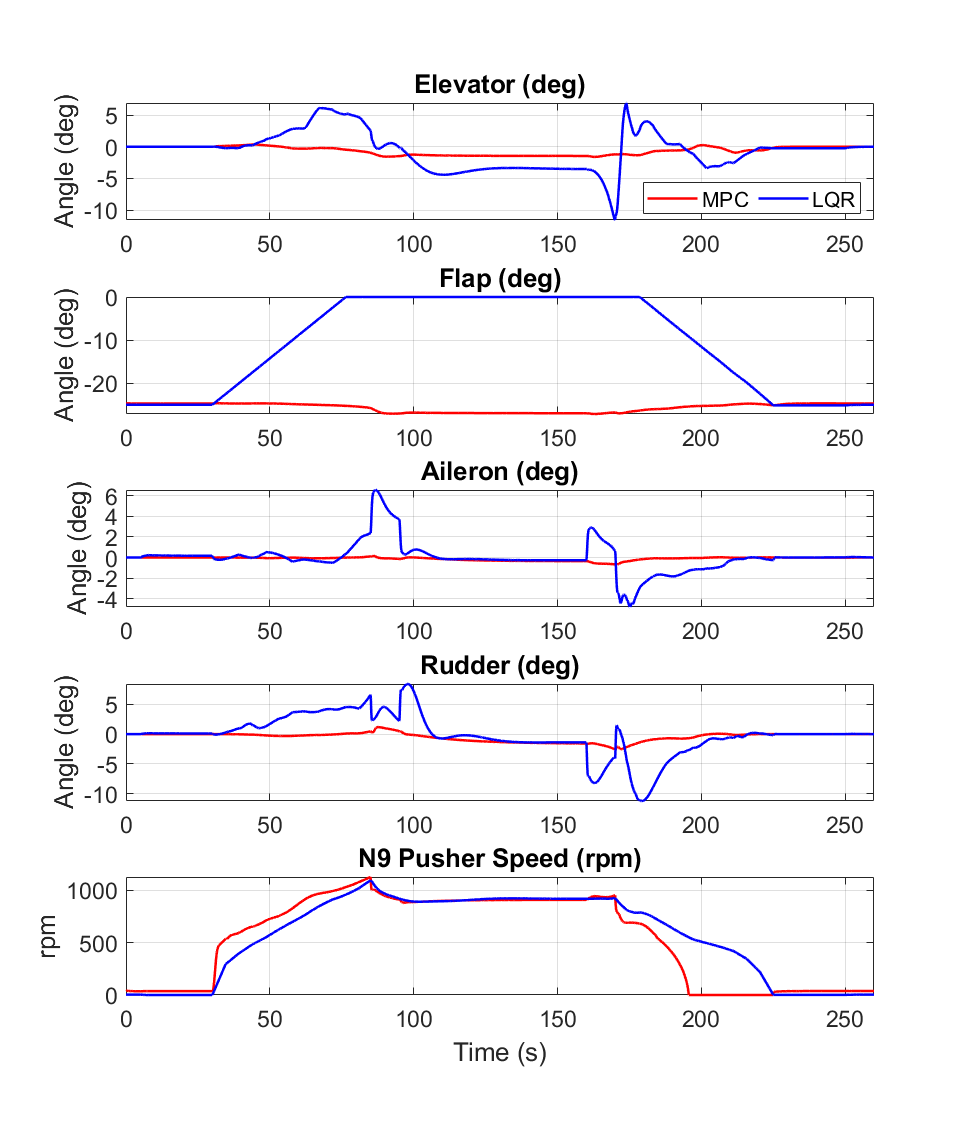}
\caption{Control surfaces and pusher propeller speed.}\label{fig:MPC_LQR_surfaces}
\end{center}\end{figure}

\begin{figure}[htbp] \begin{center}
\includegraphics[width=.8\columnwidth,trim={1cm 1cm 1.5cm 1.5cm}, clip]{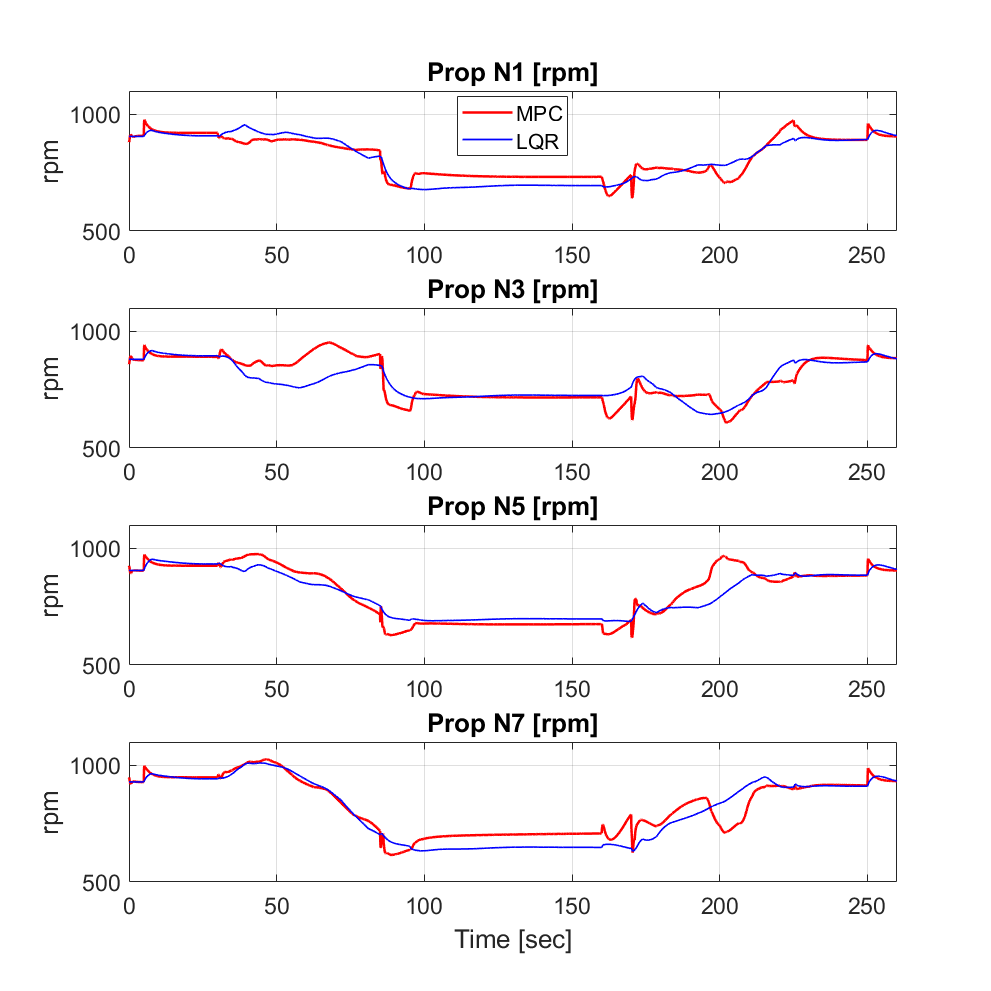}
\caption{Propeller speeds for the front-wing rotors.}
\label{fig:MPC_LQR_N1357}
\end{center}\end{figure}

\begin{figure}[htbp] \begin{center}
\includegraphics[width=.8\columnwidth,trim={1cm 1cm 1.5cm 1.5cm}, clip]{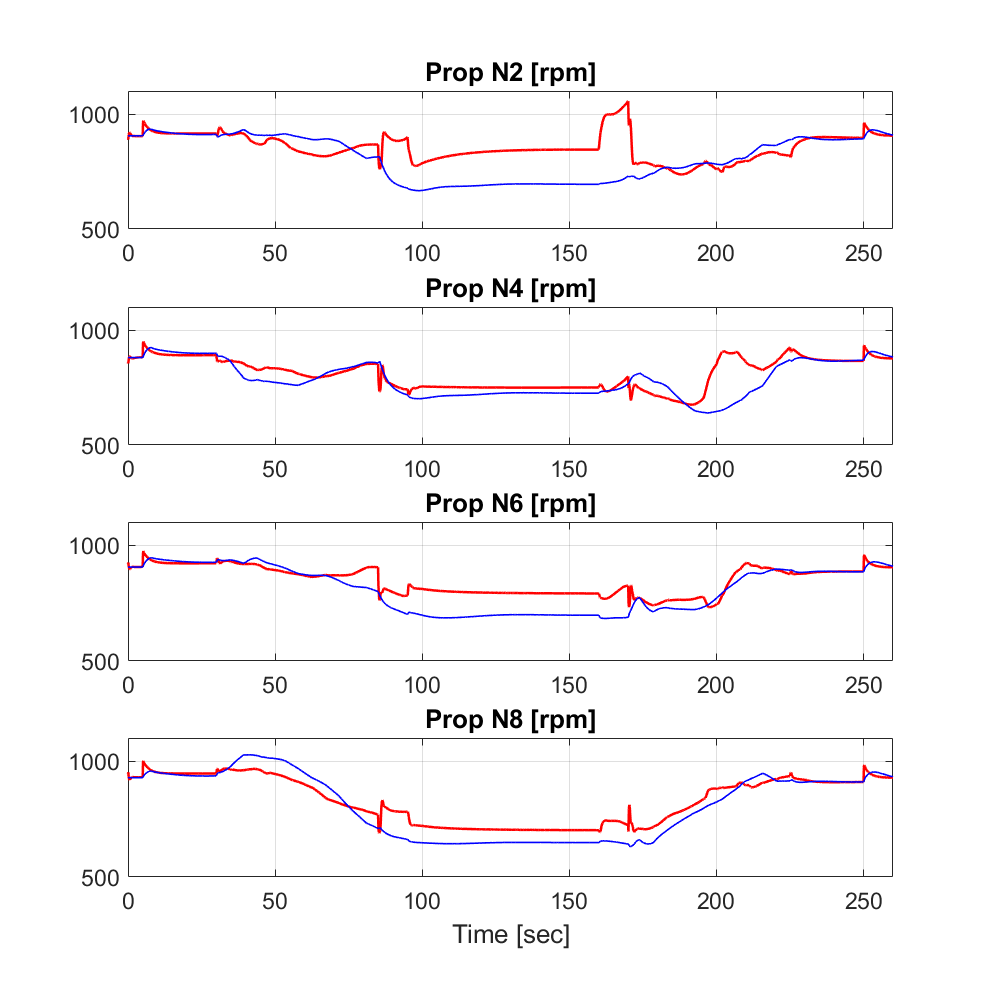}
\caption{Propeller speeds for the rear-wing rotors.}\label{fig:MPC_LQR_N2468}
\end{center}\end{figure}

The trajectory tracking results are presented in Fig.~\ref{fig:MPC_LQR_traj}–Fig.~\ref{fig:MPC_LQR_N2468}. Both controllers exhibit effective tracking performance across all flight regimes, including transitions from hover to forward flight and back to hover, while executing climb, turn, and descent maneuvers. As shown in Fig.~\ref{fig:MPC_LQR_traj}, the maximum path deviation is 55 ft for the RSLQR controller and 41 ft for the MPC controller over the course of the flight. The time histories of the translational and rotational velocities in Fig.~\ref{fig:MPC_LQR_vel} further indicate that the MPC achieves improved velocity tracking relative to RSLQR, along with stronger damping in the pitch and roll responses, which is beneficial for passenger comfort.
Fig.~\ref{fig:MPC_LQR_Euler} shows the pitch and roll angles during the flight for both the MPC and RSLQR controllers. The roll angle is used to command constant-radius turns, while the pitch angle is used to reorient the thrust vector during transition maneuvers. 

The control actuation throughout the flight is shown in Figs.~\ref{fig:MPC_LQR_surfaces}–\ref{fig:MPC_LQR_N2468}. Fig.~\ref{fig:MPC_LQR_surfaces} presents the control surface deflections, including the elevator, flap, aileron and rudder, as well as the pusher rotor speed (N9).
The LQR controller exhibits more aggressive control surface activity than the MPC during hover, ascent, descent, and turning maneuvers, where aerodynamic control surfaces are inherently less effective due to low dynamic pressure. In contrast, the MPC produces comparatively smoother control surface responses. However, during cruise conditions, the MPC still applies significant flap deflections (approximately $-27$ deg). Physically, flaps should be used primarily during the ascent and descent phases to increase lift rather than during steady cruise. Therefore, a regime-dependent penalty weight should be introduced to flap usage in the MPC formulation to discourage their use as primary control effectors during cruise flights. The pusher rotor speed (N9) becomes more active during the cruise flight (approximately 90 to 170 s) for both controllers.

Figs.~\ref{fig:MPC_LQR_N1357}-\ref{fig:MPC_LQR_N2468} show the propeller speeds (N1-N8), which provide lift in hover, thrust in forward flight, and roll, pitch, and yaw control authority across all flight regimes. The overall magnitudes are comparable between MPC and RSLQR. However, the MPC exhibits more actively modulated propeller speeds than the LQR, contributing to improved trajectory tracking performance as well as enhanced damping in pitch and roll responses.
\vspace{-0.3in}
% Translational and rotational velocities

% Roll and pitch angles.

% Control surface & N9

% N1357

% N2468

% % N1-N8 speeds
% \begin{figure*} % Remember: figure* usually prefers [t] over [b]
%     \centering    \includegraphics[width=\textwidth, trim={6.5cm 1cm 5.5cm 1.5cm}, clip]{Figs/MPC_LQR_N18.png}
%     % \vspace{-0.4 in}
%     \caption{Propeller N1-N8 speeds.}
%     \label{fig:MPC_LQR_N18}
% \end{figure*}

\subsection{Fault-tolerant control under actuator degradation}
DEP systems provide enhanced control authority and inherent fault tolerance by enabling the remaining motors to redistribute control effort following actuator failures, thereby maintaining safe operation even under degraded conditions. This capability is particularly important for autonomous systems, where contingencies must be handled without human intervention, as well as for special operations requiring simplified vehicle operation (SVO) for non-pilot users.
A key advantage of the MPC framework over LQR lies in its ability to explicitly incorporate fault tolerance in real time through reconfigurable constraint formulations that account for actuator failures, including degraded or failed DEP motors. 

\begin{figure*} \begin{center}
\includegraphics[width=\textwidth,trim={1.5cm 2cm 1.5cm 8cm}, clip]{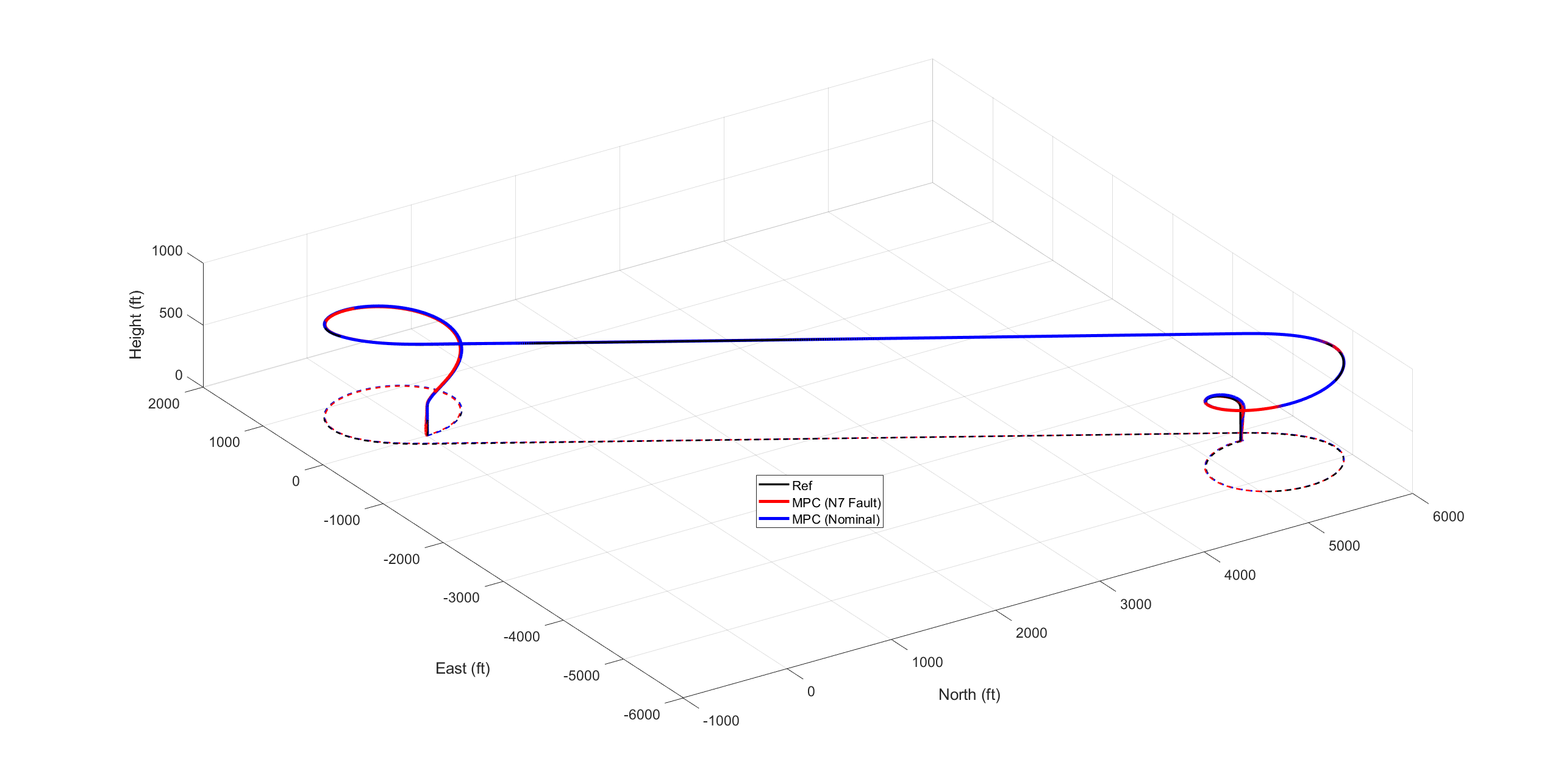}
\caption{Trajectories in fault-tolerance mode.}\label{fig:MPC_fault_traj}
\end{center}\end{figure*}

\begin{figure} \begin{center}
\vspace{-.2in}
\includegraphics[width=.9\columnwidth, trim={1cm 1cm 1cm .5cm}, clip]{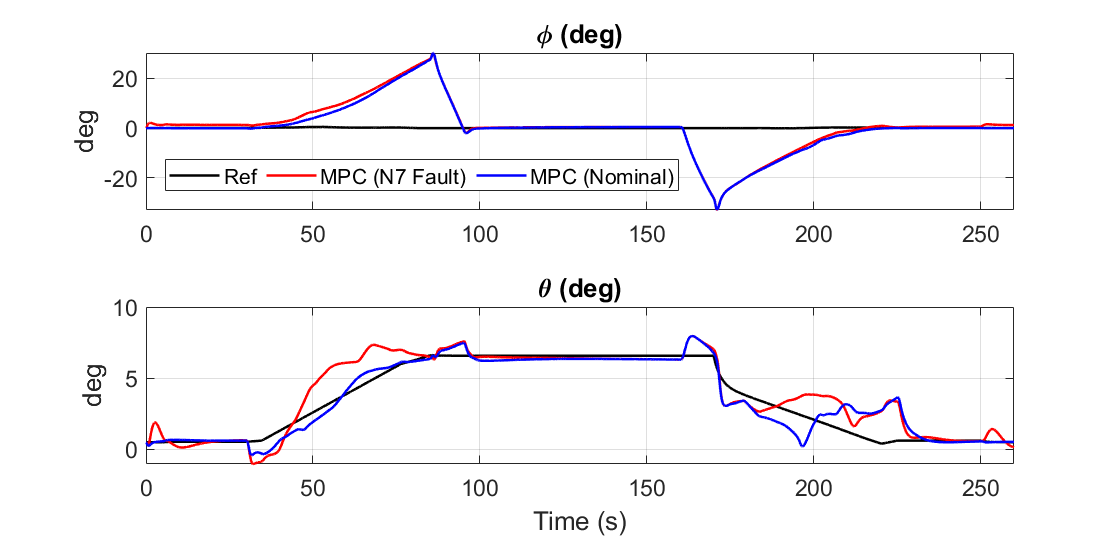}
\caption{Roll and pitch angles in fault-tolerance mode.}\label{fig:MPC_fault_Euler}
\end{center}\end{figure}

\begin{figure} % Remember: figure* usually prefers [t] over [b]
    \centering    \includegraphics[width=.9\columnwidth, trim={0cm 2cm 1cm 2cm}, clip]{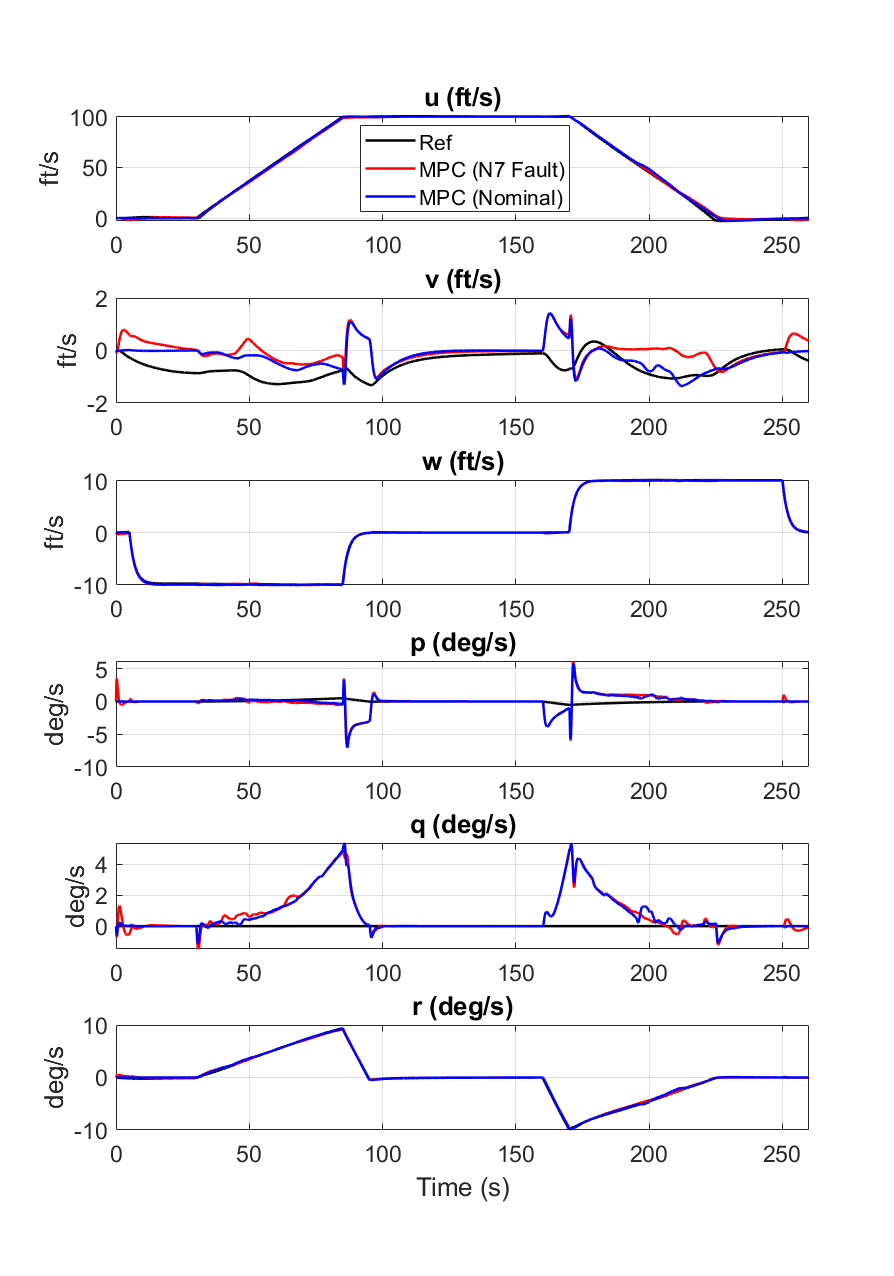}
    % \vspace{-0.5 in}
    \caption{Translational and rotational velocities in fault-tolerance mode.}
    \vspace{-.2in}
    \label{fig:MPC_fault_vel}
\end{figure}

\begin{figure} \begin{center}
\includegraphics[width=.9\columnwidth,
trim={1cm 1.5cm 1.5cm 1.5cm}, clip]{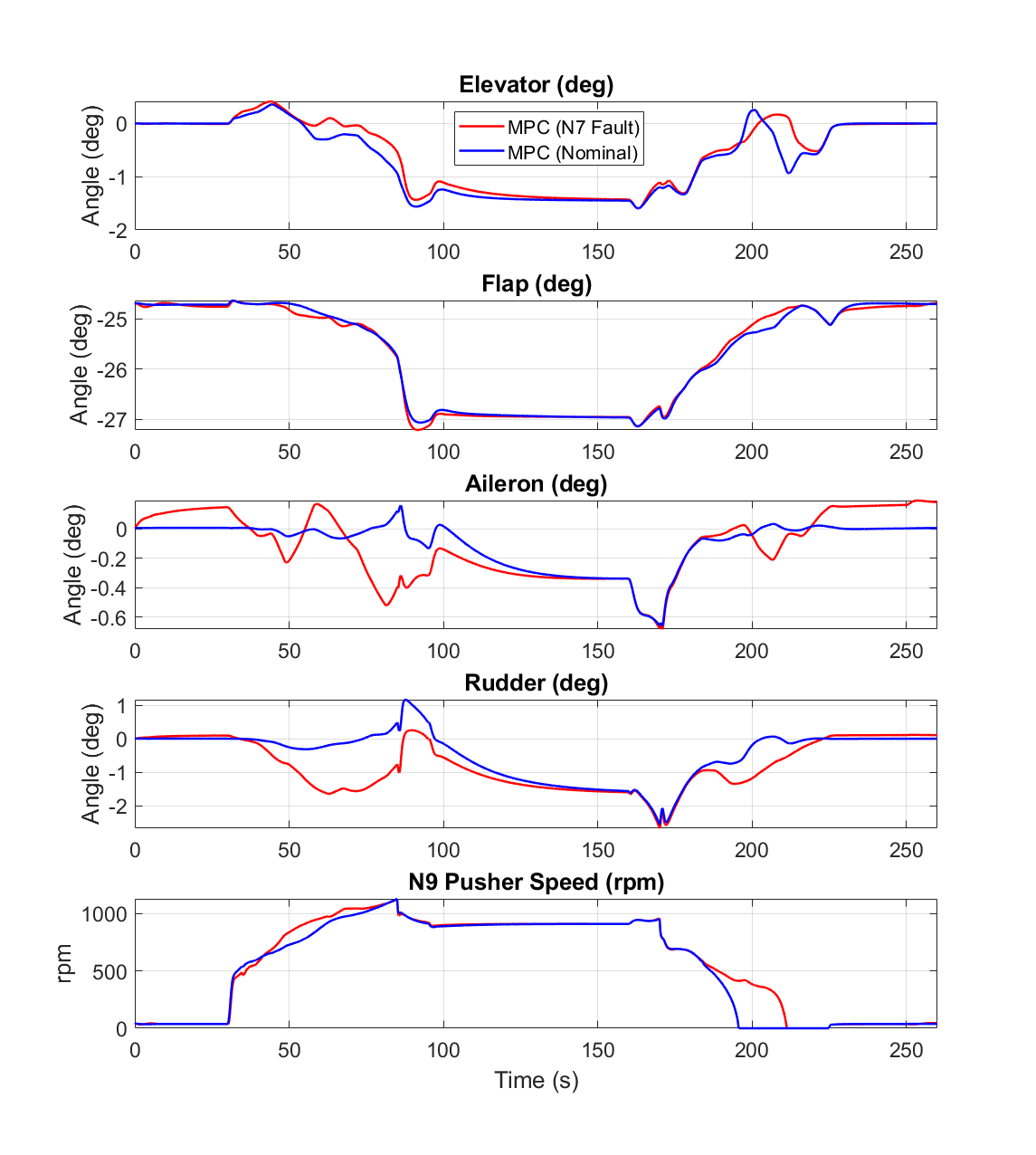}
\vspace{-.2in}
\caption{Control surfaces and pusher propeller speed in fault-tolerance mode.}\label{fig:MPC_fault_surfaces}
\end{center}\end{figure}

\begin{figure}[htbp] \begin{center}
\includegraphics[width=\columnwidth,trim={1cm 1cm 1.5cm 1.5cm}, clip]{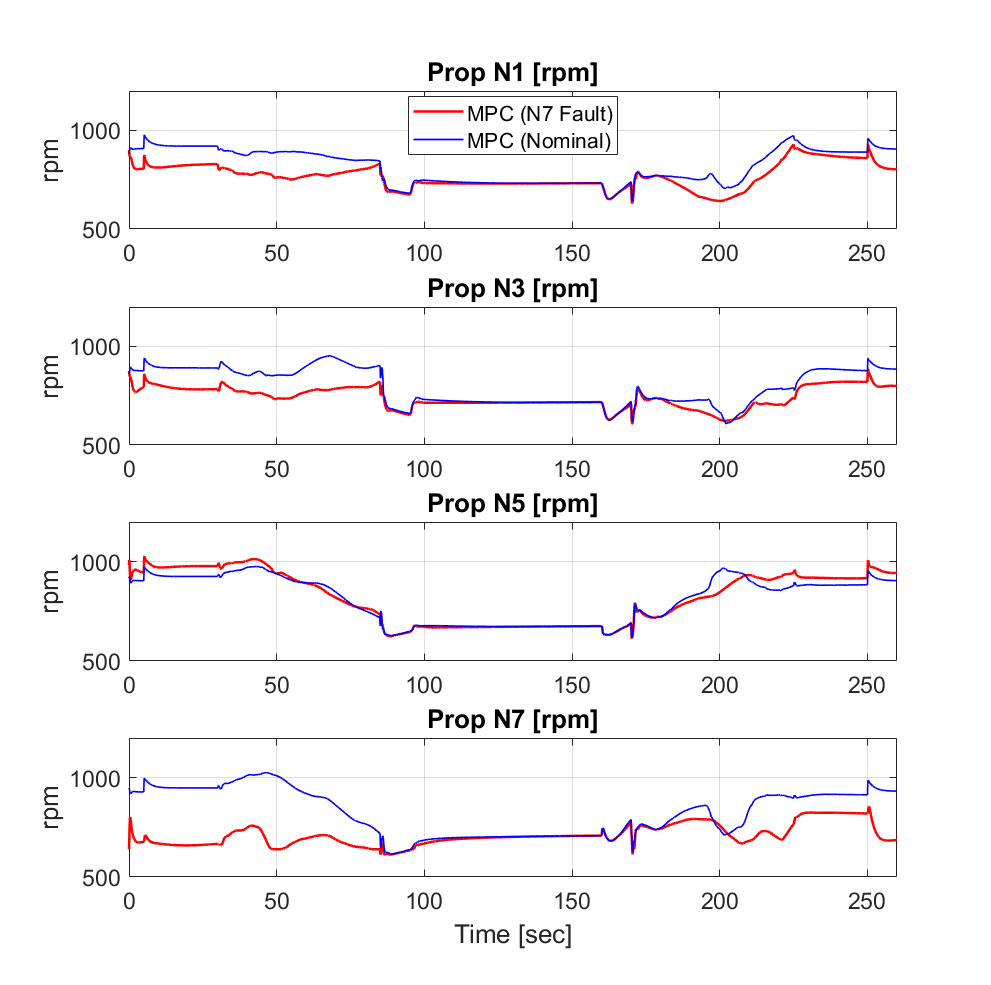}
\caption{Propeller speeds for the front-wing rotors in fault-tolerant mode (N7 with 50\% reduced authority).}
\label{fig:MPC_fault_N1357}
\end{center}\end{figure}

\begin{figure}[htbp] \begin{center}
\includegraphics[width=\columnwidth,trim={1cm 1cm 1.5cm 1.5cm}, clip]{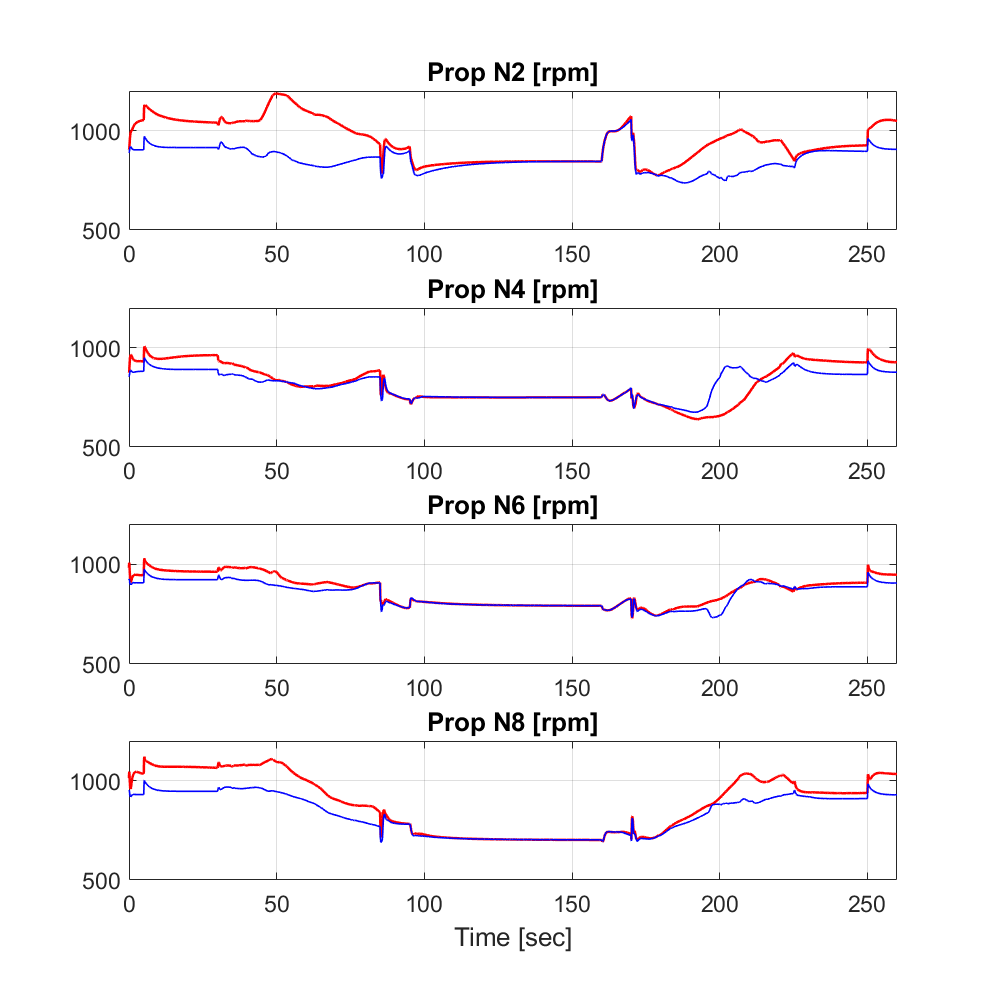}
\caption{Propeller speeds for the rear-wing rotors in fault-tolerant mode.}
\label{fig:MPC_fault_N2468}
\end{center}\end{figure}

This section demonstrates this capability by evaluating tracking performance under partial loss of control authority. In particular, Rotor N7 in Fig.~\ref{fig:LPC} is assumed to experience a 50\% reduction in control authority, representing a degraded actuator scenario. 
The trajectory tracking results are presented in Fig.~\ref{fig:MPC_fault_traj}–Fig.~\ref{fig:MPC_fault_N2468}. The MPC controller maintains effective tracking performance across all flight regimes despite the N7 fault condition. The time histories of the translational and rotational velocities in Fig.~\ref{fig:MPC_fault_vel} further indicate that the MPC preserves satisfactory velocity tracking performance relative to the nominal (fault-free) case.
Fig.~\ref{fig:MPC_fault_N1357} shows a significant reduction in the N7 propeller speed compared to the remaining propellers due to the partial loss of control authority. To compensate for this degradation, other propellers (N2, N4, N5, N6, and N8) increase their rotational speeds accordingly. Overall, the propeller commands remain well-behaved, demonstrating that the MPC effectively manages actuator degradation through constraint-aware optimal control allocation.

% Trajectory

% Roll and pitch angles.
% Translational and rotational velocities

% Control surface & N9

% N1-N8 speeds
% \begin{figure*} % Remember: figure* usually prefers [t] over [b]
%     \centering    \includegraphics[width=\textwidth, trim={6.5cm 1cm 5.5cm 1.5cm}, clip]{Figs/MPC_fault_N18.png}
%     % \vspace{-0.4 in}
%     \caption{Propeller N1-N8 speeds in the fault-tolerance mode.}
%     \label{fig:MPC_fault_N18}
% \end{figure*}

% N1357

% N2468

\section{Conclusions}

This paper presented an adaptive model predictive control framework for nonlinear eVTOL flight across the full operational envelope. The key contributions and findings are summarized as follows:

\begin{itemize}
\item An adaptive MPC architecture was developed using a LPV representation, enabling real-time adaptation of the predictive model. This formulation effectively captures the nonlinear and time-varying dynamics of the aircraft across all flight regimes, ensuring consistent performance throughout transitions. 

\item Numerical simulations demonstrated that the proposed MPC controller outperforms the baseline Robust Servomechanism Linear Quadratic Regulator in terms of trajectory tracking accuracy, control smoothness, and velocity regulation, while maintaining satisfaction of actuator and state constraints.

\item The proposed framework explicitly enforces dynamic control allocation, enabling safe operation under both nominal conditions and actuator degradation scenarios, thereby providing inherent fault-tolerant capability.
\end{itemize}

Overall, the results indicate that the proposed adaptive MPC framework is a promising candidate for next-generation autonomous eVTOL control systems operating in complex and safety-critical environments. Future work will investigate the performance of the framework under a wider range of contingency scenarios, including actuator failures and environmental disturbances, to further enhance its stability and robustness. In addition, this MPC architecture will be extended to provide a unified framework for analyzing the stability, control, and performance of different eVTOL configurations, including lift-plus-cruise, tiltrotor, and other concepts.

Author contact: Tri Ngo, tri.ngo@utdallas.edu

% \section{Appendix}

% Appendices should be used for highly specialized data, derivations, etc. They
% should be lettered (A, B, C, ...) if more than one is used. Each appendix must
% be cited in the main text. 

\bibliographystyle{ahs}
\label{sec:references}\bibliography{ahsrefs}

\end{document}